\documentclass[aps,twocolumn,showpacs,tightenlines,superscriptaddress]{revtex4}
\usepackage{amsmath}
\usepackage{amsfonts}
\usepackage{amssymb}
\usepackage{graphicx}

\begin{document}
\title{Entanglement-assisted Orientation in Space}

\author{{\v C}aslav Brukner}
\affiliation{Institut f\"ur Experimentalphysik, Universit\"at Wien, Boltzmanngasse 5,
A--1090 Wien, Austria} \affiliation{Institut f\"ur Quantenoptik und Quanteninformation,
\"Osterreichische Akademie der Wissenschaften, Boltzmanngasse 3, A--1090 Wien, Austria}
\author{Nikola Paunkovi\'{c}}
\affiliation{Fondazione I.S.I., Villa Gualino, Viale Settimio Severo 65, 10133 Torino, Italy}
\author{Terry Rudolph}
\affiliation{Department of Physics, Blackett Laboratory, Imperial College London, Prince Consort Road, London
SW7 2BW, UK}
\author{Vlatko Vedral}\affiliation{The School of Physics and Astronomy, University
of Leeds, Leeds, LS2 9JT, United Kingdom}\affiliation{Institut f\"ur Experimentalphysik,
Universit\"at Wien, Boltzmanngasse 5, A--1090 Wien, Austria}
\date{\today}

\begin{abstract}

We demonstrate that quantum entanglement can help separated individuals in making
decisions if their goal is to find each other in the absence of any communication
between them. We derive a Bell-like inequality that the efficiency of every classical
solution for our problem has to obey, and demonstrate its violation by the quantum
efficiency. This proves that no classical strategy can be more efficient than the
quantum one.

\end{abstract}

\maketitle

Quantum entanglement is a phenomenon in which two or more quantum systems have to be
described with reference to each other regardless of their spatial
separation~\cite{schroedinger}. It leads to correlations that are inconsistent with
local realism~\cite{epr} as demonstrated by violation of Bell's inequalities~\cite{bell}.
Although entanglement does not involve information transfer, it surprisingly can produce
effects over arbitrary distances as if information had been transferred. It can {\it
substitute or even eliminate any need of communication} that is classically necessary
for achieving a goal of common interest of separated parties~\cite{yao,cleve,brassard}.
Entanglement can thus reduce the communication complexity of certain problems; the key
ingredient for this is violation of local realism (quantum non-locality)~\cite{brukner0}.

Here we show that entanglement can help individuals in making decisions if their goal is
to find each other, even if there is no communication between them. This gives the
problem properties of ``pseudo-telepathy"~\cite{brassard1}. Ref.~\cite{summi} suggests a
similar mechanism to explain cooperation of insects and demonstrates an advantage of the
use of entanglement over some classical strategies. To provably demonstrate that no
classical solution of our problem exists that can achieve the efficiency of the
entanglement-based quantum solution, we derive a Bell-type inequality that every
classical efficiency has to obey and demonstrate its violation by the quantum
efficiency. In a broader context we show that quantum non-locality - one of the most
peculiar features of quantum physics - can be used to solve everyday if somewhat unusual
problems, as similarly suggested by~\cite{gameshow,wiseman}.

\begin{figure}
\centering
\includegraphics[angle=-90,width=9.2cm]{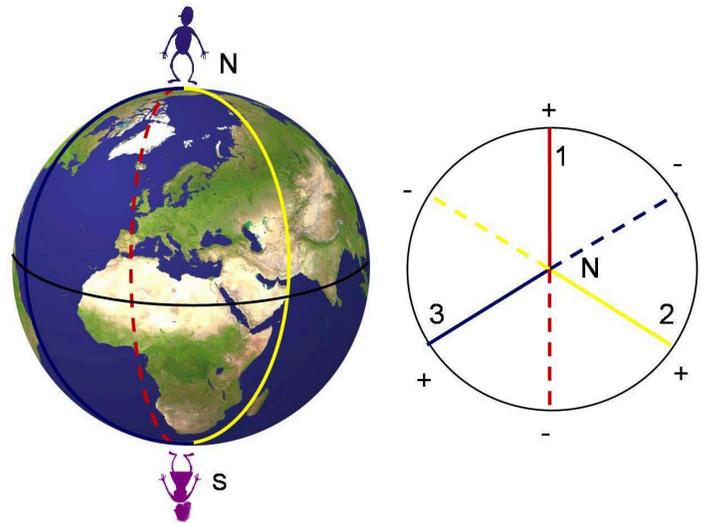}
\caption{Two partners are on the two poles of the Earth (left). From each pole there are
three paths (red 1, yellow 2 and blue 3) and for each path there are two directions (+
and $-)$ (right, view from the North pole). Which path and direction should the partners
take to find each other at the equatorial line in the lack of any communication? (see
text for details)}
\end{figure}

We now define the problem in detail. We let two partners be on the two poles of the
Earth, Alice on the North pole and Bob on the South pole. From either of the poles there
are three possible paths, red (1), blue (2) and yellow (3), each with two possible
directions, (+) and $(-)$ (Fig. 1). Positive (negative) directions of each pair of the
paths enclose an angle of $120^0$ with each other if viewed from the poles. Each path is
represented by one great circle on the globe.

With the aim to find each other, each partner chooses a path and
direction and follows it along the great circle of the globe
until she/he reaches the equatorial line. Arriving there the
partners certainly meet if they have taken the same path and the
same direction (either $\{+,+\}$ or $\{-,-\}$). Alternatively,
they may miss each other if they have taken either different
paths or different directions of the same path. In the latter
case the partners will end at two opposite points of the equator.
In the former case they will end at two points that are separated
either under angle of $60^0$, in the case they have taken
different directions (either $\{+,-\}$ or $\{-,+\}$), or under
angle $120^0$, if they have taken the same directions (either
$\{+,+\}$ or $\{-,-\}$). We will assume that both partners have a
far-reaching view with an opening angle of $60^0$ so that even if
they took different paths but opposite directions they can see and
subsequently find each other on the equator.

We assume that the partners can not communicate nor that they have agreed in advance
which path and direction to take in order to find each other. Furthermore, we assume
that the choices of paths are completely independent, random and given with equal
probabilities \footnote{Later in the text we will see that this corresponds to the
common assumption of experimenter's freedom in choosing between different possible
measurement settings in the context of Bell's inequalities. See, for example, J.S. Bell,
Free Variables and Local Causality, Dialectica \textbf{39}, 103-106 (1985).}. However, in
choosing the directions Alice and Bob are allowed to use some previously shared
classically correlated random strings or quantum entanglement. We therefore consider a
situation in which Alice and Bob share no information whatsoever about which path each
other's partner intends to take \footnote{Such a formulation of the problem invalidates
a trivial possibility for partners to share a common set of instructions of the type
''take yellow path, direction +".}.

In order to achieve their goal with the highest possible efficiency Alice and Bob should
maximize both the probability to take the same directions if they choose the same path
and the probability to take opposite directions if they choose different paths. The
overall probability of success is given by
\begin{equation}
P=\frac{1}{9} \left(\sum_{i=1}^3 P_{ii}(\mbox{same}) + \sum_{i\neq j=1}^3 P_{ij}
(\mbox{opp})\right), \label{noga}
\end{equation}
where, e.g., $P_{ij}(\mbox{opp})$ is the probability that the partners take opposite
directions if Alice chooses path $i$ and Bob $j$. The factor $1/9$ is due to the
assumption that every possible combination of the two paths taken by Alice and Bob is
equally probable. We will now give the quantum solution of the problem. It is based on
an example of quantum non-locality presented by Mermin~\cite{mermin1}.

Suppose that the partners share a maximally entangled pair of photons in the state
$|\phi^+\rangle = 1/\sqrt{2}(|H\rangle|H\rangle + |V\rangle|V\rangle)$, where
$|H\rangle$ denotes horizontal polarization and $|V\rangle$ vertical polarization of a
photon. Every partner chooses a path at random from the set $\{1,2,3\}$, independently of
each other. The choice of paths determines a choice of directions of polarization
measurements, given by the angles $\{0^0,120^0,-120^0\}$: if the taken path is $1$, the
measurement angle is $0^0$, while in case of paths $2$ and $3$, measurement angles are
$120^0$ and $-120^0$, respectively. Polarization measurements are performed on each
partner's photons. If the outcome is $H$, the partner takes direction +, and if it is
$V$ she/he takes direction $-$. In 3/9 fraction of cases the partners choose the same
measurement directions and obtain the same outcomes with certainty. In the remaining 6/9
fraction of cases they choose different measurement directions, in which case the
probability of them obtaining opposite results is 3/4. Thus, one has
$P\!=\!5/6\!\approx\! 83\%$ for the success rate in the quantum protocol.

In a classical protocol, instead of an entangled pair of photons, Alice and Bob can use
some previously distributed classically correlated random strings of variables which may
improve the success of the protocol. On the basis of these strings they could make their
decision which directions ($+$ or $-$) to choose, but the choice of the path (red,
yellow, or blue) is, as already mentioned, assumed not to be dependent on the strings.

The best classical strategy is successful in $7/9 \approx 78\%$ of the cases. This can be
achieved in a model where Alice and Bob share classically correlated variables and in
which each variable carries an internal code, determining whether the outcome $H$ or $V$
will emerge, for each of the three possible choices of paths 1, 2 and 3. Thus, the set
of all possible codes for both variables given to Alice and Bob is: $\{(HHH),(VVV),$$
(HHV), (HVH), (VHH),$ $(HVV), $$(VHV),$$ (VVH)\}$. Here the position in each code plays
the role of the measurement angle, while the corresponding value ($H$ or $V$) at that
position corresponds to a measurement outcome. Imagine that each pair of variables, one
given to Alice and the other to Bob, carries the same code, so as to have the maximal
possible probability of success in the case when both partners choose the same path:
$\frac{1}{9} \sum_{i=1}^3 P_{ii}(\mbox{same})=3/9$. For any of those pairs of codes if
the choices of paths are different, then the probability of emerging opposite outcomes
is at most $2/3$, which can be easily checked by simple inspection (the value $7/9$ is
obtained if neither of codes $\{(HHH),(VVV)\}$ is used). Thus, the efficiency is
$P\!=\!(3/9) + (6/9)(2/3)\!=\!7/9$ in this protocol.

No classical protocol can be more efficient than this, in the lack of any communication
between the two parties. This is based on the proof given below that the combination of
probabilities as given in Eq.~(\ref{noga}) satisfies a Bell inequality with 7/9 being a
local realistic bound. Thus, the probability of success of any classical protocol is
bounded, while those of quantum protocols can exceed the limit by utilizing quantum
non-locality.

We now give the proof that
\begin{equation}
\sum_{i=1}^3 P_{ii}(\mbox{same}) + \sum_{i\neq j=1}^3 P_{ij}
(\mbox{opp}) \leq 7, \label{vena}
\end{equation}
for all local realistic models. This new Bell's inequality with
three possible measurement settings per observer is a byproduct
of our analysis. Consider two observers, Alice and Bob, and allow
each of them to choose between three dichotomic observables,
determined by some local parameters denoted here $a_1$, $a_2$ and
$a_3$ for Alice and $b_1$, $b_2$ and $b_3$ for Bob. The
assumption of local realism implies the existence of three
numbers $A_1$, $A_2$ and $A_3$ for Alice and $B_1$, $B_2$ and
$B_3$ for Bob, where the numbers take values $+1$ or $-1$ and
describe the predetermined results of corresponding measurements.
In other words, for given specific numerical values, $(A_1A_2A_3)$
is a code from our previous example, while the values $+1$ and
$-1$ play the role of $H$ and $V$, respectively. In a specific
run of the experiment the correlations between two observations
can be represented by the product of the type $A_i B_j$. The
correlation function is then the average over many runs of the
experiment $E(a_i,b_j)=\langle A_i B_j \rangle$.

The following combination of the predetermined results has a
maximal value as given by:
\begin{eqnarray}
\mbox{max} \{A_1(B_1-B_2-B_3) &+& A_2(B_2-B_1-B_3) \label{nocas}\\
&+& A_3(B_3-B_1-B_2)\}=5  \nonumber
\end{eqnarray}
After averaging this expression over the ensemble of the runs of the experiment, one
obtains the following Bell inequality:
\begin{equation}
| \sum_{i=1}^{3} E(a_i,b_i) - \sum_{i\neq j}^{3} E(a_i,b_j) | \leq
5.
\end{equation}
Using the connection between correlation functions and probabilities
\footnote{$E(a_i,b_j)=\langle A_i B_j \rangle=\sum_{A_i, B_j \in \{ -1,1\}} P(A_i,
B_j)A_i B_j=P_{ij}(\mbox{same})-P_{ij}(\mbox{opp})=2 P_{ij}(\mbox{same}) -1=1-2
P_{ij}(\mbox{opp})$, as $P_{ij}(\mbox{same})+P_{ij}(\mbox{opp})=1$. Here $P(A_i, B_j)$
is probability that Alice obtains $A_i$ and Bob $B_j$, for the choices of paths $i$ and
$j$, respectively.}, $E(a_i,b_i)\!=\!2 P_{ii}(\mbox{same}) -1$ and $E(a_i,b_j)\!=\!1-2
P_{ij}(\mbox{opp})$ when $i \neq j$, one finally obtains inequality~(\ref{vena}). Its
local realistic limit demonstrates that the efficiency of the particular classical
solution discussed above is the optimal one. The quantum solution is based on violation
of the inequality by a factor $7.5$ with the quantum entangled state $|\phi^+\rangle$.

Our problem can be seen as an instance studied in the field of communication complexity.
Typically, in communication complexity problems two spatially separated parties receive
local input data, e.g. one party receives number $i$ and the other $j$. Their goal is to
compute a given function $f(i,j)$. In one class of these problems only a restricted
amount of communication between the parties is allowed, so that in general they can not
arrive at the correct value of the function with certainty. While an error is allowed,
the parties try to compute the function correctly with as high probability as possible.

In our example no communication between Alice and Bob is allowed. The input data $i$ and
$j$ correspond to Alice's and Bob's choices of paths, respectively. Function $f(i,j)$ is
defined in the following way: $f(i,i)=1$ and $f(i,j)\!=\!-1$ if $i \!\neq \!j$. On the
basis of $i$ and $j$ Alice and Bob produce new local data $A$ and $B$ (obtained as
results $+1$ or $-1$ of local measurements of polarization along directions $i$ and $j$
in quantum protocol) that define the directions of their movements along paths $i$ and
$j$, respectively. Only if $f(x,y)\!=\!A \cdot B$ they will find each other (in which
case they also compute the function correctly).

In conclusion, we have demonstrated that quantum entanglement can help separated
individuals to orient themselves on a sphere if their aim is to find each other, even in
the absence of any communication between them. In future it will be interesting to
investigate whether there are further problems in life science, economics or every-day
situations whose solutions favor quantum correlations over classical ones.

{\v C}.B. has been supported by the Austrian Science Foundation (FWF) Project SFB 1506
and the European Commission, Contract No. IST-2001-38864 RAMBOQ. N.P. is funded by the
European Commission, Contract No. IST-2001-39215 TOPQIP. T.R. is supported by the
Engineering and Physical Sciences Research Council.

The authors gratefully acknowledge discussions with Karl Svozil and  Johann Summhammer
and hospitality of Oxford's ''The Three Goats' Heads" pub and Vienna coffee house
''Pr\"{u}ckel".

\end{document}